\documentclass[prl,twocolumn,showpacs,superscriptaddress]{revtex4}

\usepackage{graphicx}

\begin{document}

\title{Quantum mechanics in metric space: wave functions and 
their densities}

\author{I. D'Amico}\email{ida500@york.ac.uk}
\affiliation{Department of Physics, University of York, York YO10 5DD,
United Kingdom}

\author{J. P. Coe}
\affiliation{Department of Physics, University of York, York YO10 5DD, 
United Kingdom}

\author{V. V. Fran\c{c}a}
\affiliation{Physikalisches Institut, Albert-Ludwigs Universitat, Hermann-Herder-Str. 3, Freiburg, Germany}
\affiliation{Capes Foundation, Ministry of Education of Brazil, Caixa Postal 250, Brasilia, 70040-20, Brazil}

\author{K. Capelle}
\affiliation{Centro de Ci\^encias Naturais e Humanas,
Universidade Federal do ABC, Santo Andr\'e, 09210-170 S\~ao Paulo,
Brazil}

\date{\today}

\begin{abstract}
Hilbert space combines the properties of two fundamentally different types of 
mathematical spaces: vector space and metric space. While the vector-space 
aspects of Hilbert space, such as formation of linear combinations of state 
vectors, are routinely used in quantum mechanics, the metric-space aspects
of Hilbert space are much less exploited. Here we show that a suitable metric 
stratifies Fock space into concentric spheres. Maximum and minimum distances between wave 
functions are derived and geometrically interpreted in terms of this metric.
Unlike the usual Hilbert-space analysis, our results apply also to the 
reduced space of only ground-state wave functions and to that of 
particle densities, each of which forms a metric space but {\em not} 
a Hilbert space. The Hohenberg-Kohn mapping between densities and ground-state 
wave functions, which is highly complex and nonlocal in coordinate description,
is found, for three different model systems, to be very simple in metric space,
where it is represented by a monotonic mapping of vicinities onto vicinities. 
Surprisingly, it is also found to be nearly linear over a wide range of 
parameters.
\end{abstract}



\newcommand{\be}{\begin{equation}}
\newcommand{\ee}{\end{equation}}
\newcommand{\bea}{\begin{eqnarray}}
\newcommand{\eea}{\end{eqnarray}}
\newcommand{\bi}{\bibitem}
\newcommand{\la}{\langle}
\newcommand{\ra}{\rangle}
\newcommand{\ua}{\uparrow}
\newcommand{\da}{\downarrow}
\renewcommand{\r}{({\bf r})}
\newcommand{\rp}{({\bf r'})}
\newcommand{\eps}{\epsilon}
\newcommand{\bfr}{{\bf r}}

\maketitle

It is a fundamental tenet of quantum mechanics that the space of the proper
wave functions of an $N$-particle quantum system is a Hilbert space: a
complete vector space of (complex) functions among which a scalar product 
is defined \cite{vonneumann}. Mathematically, a Hilbert space combines the 
properties of two fundamentally different types of spaces: a vector space 
and a metric space. In short, a vector space is one in which linear 
combinations of its elements can be formed, while a metric space is one in 
which to each two elements one can assign a distance.

The vector-space aspects of Hilbert space are widely known, and routinely
exploited in quantum mechanics. Wave functions are added,
and multiplied by real or complex numbers, to form new wave functions,
and the degree of similarity of two wave functions is measured by their
overlap, which in turn is obtained from their scalar product. Much less
exploited in quantum mechanics are the metric-space aspects of Hilbert
space. In the present Letter we explore the space of quantum-mechanical 
wave functions from the point of view of metric spaces, in which the 
similarity of two wave functions is characterized by a suitable 
metric (a measure of distance \cite{footnote1}).

Our analysis reveals several deep and useful properties of wave functions 
that are obfuscated by the more common analysis in terms of vector spaces and 
scalar products. Notably, ground-state (GS) wave functions on their own do not 
form a Hilbert space, while they still {\em do} form a metric space. Similarly, 
particle densities do not form a Hilbert space but another metric 
space. (These affirmations are explained and proven below). The 
characterization of the set of GS wave functions and densities as metric 
spaces also provides a new perspective on the Hohenberg-Kohn one-to-one
 mapping between  GS wave functions  and their densities.  

This Letter has three parts. In the first part we describe the geometry of 
the space of wave functions and their densities from the view point of metric 
space. In the second part we focus on the special case of GS wave functions.
In the last part we present results from numerical investigations of three 
model systems, illustrating and corroborating our analytical considerations.

{\em 1. Geometry of the space of wave functions and their densities.} 
We adopt the convention that the $N$-particle wave function is normalized to the total
particle number \cite{footnote3} and define the distance between any two $N$-particle wave
functions as
\bea 
& & D_\psi(\psi_1,\psi_2) = \min_\phi \tilde{D}_\psi(\psi_1,\psi_2)
\label{dist_psi0} \\
&=& \min_\phi 
\left[\int |\psi_1(x_1,..x_N)-\psi_2(x_1,..x_N)|^2 dx_1..dx_N\right]^{1/2}
\label{dist_psi1}
\\
&=& \left[\int(|\psi_1|^2+|\psi_2|^2)dx_1..dx_N  
-2\left|\int\psi_1^*\psi_2 dx_1..dx_N\right|\right]^{1/2}.
\label{dist_psi3}
\eea
where the argument $x$ represents 
spatial and spin coordinates, in any dimensionality, and the phase $\phi$
is defined through 
$\int\psi_1^*\psi_2 dx_1..dx_N=e^{i \phi}|\int\psi_1^*\psi_2 dx_1..dx_N|$
\cite{footnote2}.

Such a distance has been used for wave functions previously \cite{Longpre} and can be identified as a special case of the Bures distance \cite{Bures} applied to pure states \cite{Hubner}. In particular, the triangular inequality
\be
D_\psi(\psi_1, \psi_2)\le D_\psi(\psi_1, \psi_3)+D_\psi(\psi_3, \psi_2)\label{tri-psi}
\ee
is satisfied, and the space of all wave functions, with this
measure of distance, becomes a metric space \cite{footnote1}. 

Wave functions differing only by a constant phase are assigned distance zero 
by the metric (\ref{dist_psi1}). Without the minimization over $\phi$, the 
distance between two wave functions differing only by a constant phase,
$\psi_1$ and $\psi_2=\psi_1 e^{i\phi}$, would be 
$\tilde{D}_\psi(\psi_1,\psi_2)=2 \sqrt{N} |\sin{\phi/2}|$, 
which also satisfies the mathematical requirements for a metric, but as a 
function of $\phi$ takes on any value from 0 to $2\sqrt{N}$.
The alternative measure of distance $\tilde{D}_\psi$ thus discriminates 
between different gauge copies of the same wave function, which is unphysical.

The density of an $N$-particle wave function is 
\be
\rho(x)= \int\, |\psi(x,x_2,..x_N)|^2 dx_2..dx_N,
\label{densdef}
\ee
where $\int\rho(x)dx=N$. 
We define the distance between any two densities as
\be
D_\rho(\rho_1,\rho_2)=\int \sqrt{|\rho_1(x)|^2+|\rho_2(x)|^2-2\rho_1(x)\rho_2(x)}dx.
\label{dist_rho2}
\ee
which satisfies the  triangular inequality
\be
D_\rho(\rho_1, \rho_2)\le D_\rho(\rho_1, \rho_3)+D_\rho(\rho_3, \rho_2).\label{tri-rho}
\ee
With this definition, the space of all densities forms a metric space \cite{footnote1}, although
densities do not form a vector space, and much less a Hilbert space.

\begin{figure}
\includegraphics[width=7.5cm]{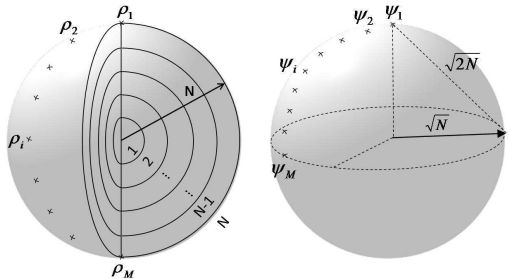}
\caption{Sketch of the metric spaces for particle densities (left) and wave functions (right). The onion-shell like geometry is explicitly displayed for the particle densities. One set of particle densities and one of wave functions are presented: their distance from the reference state -- at the north pole -- increases with their labeling index, up to maximum distance. 
}
\label{fig1}
\end{figure}

Definitions (\ref{dist_psi1}) and (\ref{dist_rho2}) are derived in the
standard way from
the characteristic norm of the quantities of interest  and determine the geometry 
of wave function and density spaces, without making use of Hilbert-space 
concepts such as scalar products or linear combinations. From 
Eq.~(\ref{dist_rho2}) it follows that all densities integrating to the same 
fixed number $N$ of particles lie in density space on a sphere of radius $N$, 
centered at the zero-density function $\rho^{(0)}(x) \equiv 0$, because 
$D_\rho(\rho,\rho^{(0)})= N$. Similarly, from Eq.~(\ref{dist_psi1}) it follows 
that all $N$-particle wave functions lie in wave function space on spheres of 
radius $\sqrt{N}$ centered at the zero-wave-function $\psi^{(0)} \equiv 0$. 
Both the space of all densities and the space of all wave functions can thus 
be visualized as concentric spheres, {\em i.e.} have an onion-shell-like 
geometry, illustrated in Fig.~\ref{fig1}. The direct sum of all $N$-particle 
Hilbert spaces is frequently denoted Fock space. From the metric point of view,
Fock space is thus stratified into an infinite number of concentric spheres, 
each of which represents an $N$-particle metric space.

An upper bound to the maximum distance between two $N$-particle densities can 
be deduced from the normalization together with the triangular inequality (\ref{tri-rho}), 
by taking 
$\rho_3 = \rho^{(0)}$:
\be  
D_\rho(\rho_1,\rho_2)\le D_\rho(\rho_1,\rho^{(0)})+D_\rho(\rho^{(0)},\rho_2)
=2N,
\ee
which is in agreement with the radius of the sphere being $N$.
This upper bound is attained in the limiting case of non-overlapping densities.
This can be seen from
Eq.~(\ref{dist_rho2}) by noting that $\rho_1(x)\rho_2(x) \ge 0$,
and that the maximum of the right-hand side is obtained when 
$\rho_1(x)\rho_2(x) \equiv 0$, so that $D_\rho^{max}=\int\,\rho_1(x) dx + \int\,\rho_2(x) dx =2N$.
Maximally distant densities are thus found to be non-overlapping densities.
Interestingly, for wave functions the situation is not that simple.

We can deduce a similar upper bound to the distance between two $N$-particle 
wave functions from the  triangular inequality (\ref{tri-psi}), taking $\psi_3 = \psi^{(0)}$:
\be
D_\psi(\psi_1, \psi_2) \le 
D_\psi(\psi_1,\psi^{(0)}) + D_\psi(\psi^{(0)},\psi_2) = 2\sqrt{N},
\label{triang_ineq_psi}\ee
which is in agreement with the radius of the sphere being $\sqrt{N}$.
However from Eq.~(\ref{dist_psi3}) it is clear that the maximum distance 
between two $N$-particle wave functions is reached for non-overlapping 
functions, and is $\sqrt{2N}$. The upper bound coming from the triangular 
inequality  is thus not attained if distances between wave functions are 
measured by $D_\psi$. If distances are measured by the alternative metric
$\tilde{D}_\psi$ (which assigns nonzero distance to wave functions differing
by a constant phase), on the other hand, the maximum distance of 
$2\sqrt{N}$ is reached for $\phi=\pi$, {\em i.e.} for $\psi$ and $-\psi$.

This situation has a simple geometric interpretation, illustrated in 
Fig.~\ref{fig1}. Start with some 
arbitrary wave function $\psi$ on the $N$-particle sphere and call it the 
north pole. The distance $D_\psi$ from the north pole to any wave function 
that does not overlap with it is the locus of points on the sphere at linear 
distance $\sqrt{2N}$. Since the diameter of the sphere is $2\sqrt{N}$ and 
$\cos\alpha=\sqrt{2N}/(2\sqrt{N})$ implies $\alpha=\pi/4$, this locus is the 
equator. 
According to $\tilde{D}_\psi$, on the other hand, the wave function that 
is maximally distant from the north pole is just $-\psi$, {\em i.e.} the
south pole \cite{footnote4}. The geometrically intuitive interpretation of maximum distance
from the north pole is thus only recovered if gauge copies of the initial wave 
function are considered to be distinct wave functions, which is unphysical. 
The same 
characterization of maximally distant wave functions according to $D_\psi$
(at the equator) and maximally distant wave functions according to 
$\tilde{D}_\psi$ (at the south pole) holds for any arbitrary wave function
taken to be the north pole, and repeats itself on any of the infinite number 
of concentric spheres in wave function metric space \cite{footnote5}.  

The minimal distance among wave functions (or densities) with same $N$ 
is obviously zero. According to Eqs.~(\ref{dist_psi1}) and (\ref{dist_rho2}) 
distances between wave functions or densities with different numbers of 
particles are not defined in general. We can, however, define a `minimal' 
distance as the difference between the radii of the corresponding 
spheres, {\em i.e.} $D_\rho^{min}(\rho^{N},\rho^{N'}):=|N-N'|$ and
$D_\psi^{min}(\psi^{N},\psi^{N'}):=|\sqrt{N}-\sqrt{N'}|$. In 
particular, the `minimal' distance between an $N$ and an $N+1$ particle 
density is 1 for all $N$, while that between the corresponding wave functions 
is $\sqrt{N+1}-\sqrt{N}$, which goes to zero for large $N$. In this sense,
{\em densities provide a better resolution for quantifying distances between 
large systems than wave functions}, which is a very interesting property in 
view of the other advantages densities have over wave functions for large 
systems \cite{kohnrmp}. 

{\em 2. Geometry of the space of ground-state wave functions.}
All of the above applies to any proper wave function of a nonrelativistic $N$-body quantum system. In the remainder of this
Letter we take a closer look at what may be the most important sub-set of
wave functions, namely those describing ground-states.

One key property of a vector space is that it is closed with respect to any 
possible linear combination, {\em i.e.}, its elements can be summed, and 
multiplied by scalars, and the result of these operations is still
an element of the same space. This property is fundamental for quantum 
mechanics, where linear combinations of wave functions abound.
We note, however, that {\em ground-state}
wave functions on their own do not satisfy this requirement: the sum of
two GS wave functions is guaranteed to be another wave function,
but not necessarily another GS wave function. Hence, the
set of all GS wave functions {\em is not a vector space}, much 
less a Hilbert space. Nevertheless, it is still a metric space, and all of
our above considerations apply just as well when restricted to ground states even if
degenerate.

\begin{figure}
\includegraphics[width=7cm]{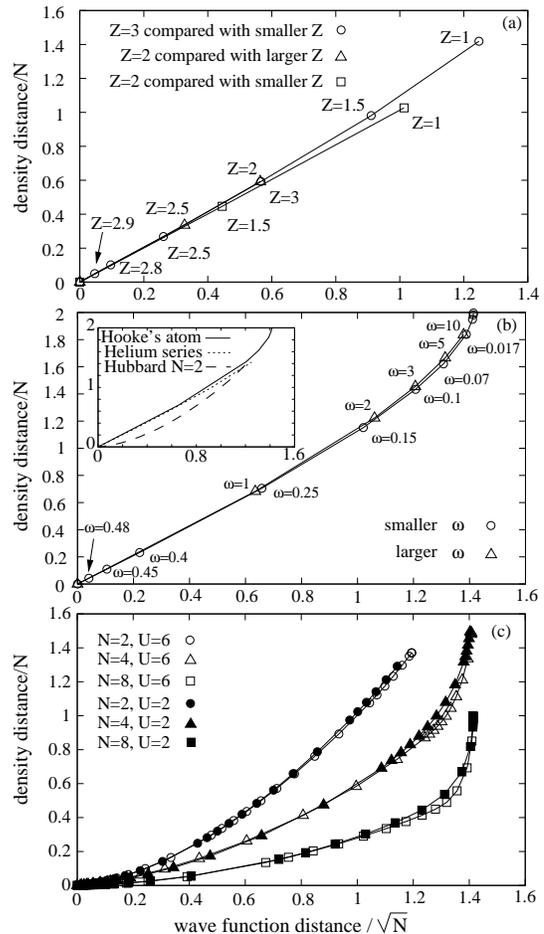} 
\caption{Density distance vs. wave function distance. Panel (a): Helium-like atoms. The reference systems have $Z=3$ (circles, decreasing $Z$) 
and $Z=2$ (triangles for increasing $Z$ and squares for decreasing $Z$). Panel (b): Hooke's atom. The
reference system has $\omega=0.5$ (triangles for increasing $\omega$ and circles 
for decreasing  $\omega$). Panel (c): Hubbard model. The reference system has $\omega=4$, 
and is compared with smaller $\omega$ systems with 8 sites and different particle numbers (circles for $N=2$, triangles for $N=4$ and squares for $N=8$) and interactions (filled symbols for $U=2$ and unfilled for $U=6$).} 
\label{fig2}
\end{figure}

The metric-space viewpoint also sheds new light on the famous mapping between 
GS wave functions and their densities, whose existence is the content of the 
Hohenberg-Kohn (HK) theorem \cite{hk}. Since GS wave functions by themselves 
do not form a Hilbert space, this mapping can be considered as one
between mere sets, {\em i.e.} structureless collections of objects. 
We have just seen, however, that both GS wave functions and densities 
form spaces with a metric structure. This metric structure can be used
to further analyze the density--wave function mapping.

By Eq.~(\ref{densdef}), a wave function uniquely determines its density.
The inverse is much less trivial, and actually is the content of the HK 
theorem: to each GS density corresponds a unique GS wave 
function in Eq.~(\ref{densdef}). 
Since any metric $D(x,y)$ satisfies $D(x,y)=0 \Leftrightarrow x=y$, the HK 
theorem implies that GS wave functions with nonzero distance 
are mapped onto densities with nonzero distance.

Due to the enormous complexity of the HK $\psi-\rho$ mapping (which is highly 
nonlinear and nonlocal in coordinate description) and the endless variety of possible functions it 
connects, analytical results on the geometry and topology of the mapping 
are very hard to obtain. However, further progress can be made by numerical 
calculations for model systems.

{\em 3. Numerical calculations.}
In this last part of the paper we consider three different nontrivial 
model systems, taken from different realms of quantum mechanics: the
one-dimensional Hubbard model, the Helium isoelectronic series, and a 
parabolically confined
two-electron system (sometimes known as Hooke's atom). For each of these
we calculate numerically highly precise or analytical GS wave functions and
densities, and investigate how a change from one wave function to 
another affects the corresponding densities. To this end we adopt one or few
states of each system as a reference state and then vary some system 
parameters (atomic number $Z$ for the He atom, the frequency $\omega$ of the
confining potential for parabolic confinement, and the frequency $\omega$ of 
the confining potential for different particle numbers $N$ and interactions $U$
for the Hubbard model). This variation produces paths on the respective metric spheres, 
{\em i.e.}, families of GS wave functions and densities, the distance of which 
from the reference states we quantify by the respective metrics. We then 
calculate $D_\rho$ as a function of $D_\psi$, as the system parameters are 
varied (see Fig.~\ref{fig2}). 

The HK theorem guarantees that the graph of $D_\rho$ as a function of $D_\psi$
starts with positive slope at the origin (where $D_\psi=D_\rho=0$) and then never reaches the horizontal ($D_\psi$) axis again. For nonzero $D_\psi$, the curves display
various additional remarkable features that are not automatic consequences of the HK
theorem.

First, in all investigated systems the initial slope is $\leq$ 45 degrees,
because the density is an integrated functional of the wave function, and as such $D_\rho$
should be affected at most as much as $D_\psi$ by a small change in the wave function. Interestingly,
the slope remains positive for the entire range of $D_\psi$, i.e., all curves increase
monotonically. At least {\it for these three model systems, the HK mapping thus consistently maps 
nearby densities onto nearby wave functions, and distant densities onto distant wave functions.}

Second, after starting at the origin, the density distance does not
only increase monotonically as a function of the   wave function distance, but even
almost linearly. In this sense, the HK mapping is as simple as it could be: 
an increase in the distance between two densities is followed by a proportional
increase in the distance between the two wave functions associated with the 
densities via the HK theorem. This near-linearity has apparently not been 
noted, and much less exploited, in the literature on density-functional theory 
or on quantum mechanics.

Third, this linearity persists up to values of $D_\psi$ that are close to
the maximum possible distance of $\sqrt{2N}$, derived above. Only for 
wave functions that are close to maximally distant ({\em i.e.} nearly
non-overlapping, according to our above analysis) do the corresponding
density distances depart from linearity and grow rapidly towards their own
maximum value $D_\rho^{max}=2N$. In this region a small increase in
the distance between two wave functions can produce a large increase 
in the distance between the two densities. Densities thus appear to be 
a rather suitable diagnostic tool for distinguishing wave functions, 
although much explicit information contained in the wave functions is integrated out 
when calculating the density. 

Finally, we note that curves for $N=2$ in different model systems can be almost 
superimposed onto each other (see inset of Fig.~\ref{fig2}-b), which hints at universality across different systems 
in the shape of the mapping. For larger $N$ a similar universality is also suggested by the Hubbard model (Fig.~\ref{fig2}-c),
since the mapping is essentially the same for different interactions.

The metric viewpoint can be applied anywhere in quantum mechanics. It could help, {\it e.g.}, in testing approximate density functionals, which should reproduce the near-linearity of the HK mapping. In variational calculations, the existence of a suitable measure of distance between wave functions may complement work such as that of Ref.~\cite{SnajdrJChemPhys2000} and help choosing better trial functions. Another use is in the development of order($N$) methods for electronic-structure calculations, as our results show that, for large $N$, densities provide better resolution for distinguishing physical systems than wave functions.  Our findings on the distance between wave functions may complement the use of the
fidelity\cite{fidelity} in quantum information theory and in the study of quantum phase transitions.  In conclusion, we emphasize that the metric and the vector space viewpoints are complementary, and that only together do they exhaust the full 
richness of Hilbert space. 

{\bf Acknowledgments} 
VVF and KC were supported by Brazilian funding agencies FAPESP, CNPq and CAPES.

\end{document}